# Characterization of the norepinephrine-activation of adenylate cyclase suggests a role in memory affirmation pathways

## Overexposure to epinephrine inactivates adenylate-cyclase, a causal pathway for stress-pathologies

ALFRED BENNUN

*Graduate Schools of Rutgers, the State University of New Jersey (R)*

## Abstract

Incubation with noradrenaline (norepinephrine) of isolated membranes of rat's brain corpus striatum and cortex, showed that ionic-magnesium ($Mg^{2+}$) is required for the neurotransmitter activatory response of Adenylate Cyclase [ATP pyrophosphate-lyase (cyclizing), (EC 4.6.1.1)], AC.

An $Mg^{2+}$-dependent response to the activatory effects of adrenaline, and subsequent inhibition by calcium, suggest capability for a turnover, associated with cyclic changes in membrane potential and participation in a short term-memory pathway.

In the cell, the neurotransmitter by activating AC generates intracellular cyclic AMP. Calcium entrance in the cell inhibits the enzyme. The increment of cyclic AMP activates kinaseA and their protein phosphorylating activity, allowing a long term memory pathway. Hence, consolidating neuronal circuits, related to emotional learning and memory affirmation.

The activatory effect relates to an enzyme-noradrenaline complex which may participate on the physiology of the fight or flight response, by prolonged exposure. However, the persistence of an unstable enzyme complex turns the enzyme inactive. Effect concordant, with the observation that prolonged exposure to adrenaline, participate in the etiology of stress triggered pathologies.

At the cell physiological level AC responsiveness to hormones could be modulated by the concentration of Chelating Metabolites. These ones produce the release of free $ATP^{4-}$, a negative modulator of AC and the $Mg^{2+}$ activated insulin receptor tyrosine kinase (IRTK). Thus, allowing an integration of the hormonal response of both enzymes by ionic controls. This effect could supersede the metabolic feedback control by energy-charge. Accordingly, maximum hormonal response of both enzymes, to high $Mg^{2+}$ and low free $ATP^{4-}$, allows a correlation with the known effects of low caloric intake increasing average life expectancy.

*Keywords: Enzyme Activation, Adenylate Cyclase, Memory, Epinephrine [Adrenaline], Stress, Longevity, Psychosomatic illness, Ionic control over Energy Charge.*

## Materials and methods

Noradrenaline, caffeine, oestradiol benzoate and ATP with GTP (obtained from horse's muscles, containing 96% of ATP, 2% of GTP and minimum quantity of others nucleotides), ATP without GTP from Sigma Chemical Co. (St. Louis, MO, U.S.A.) ;

EDTA (disodium salt) and Barbital (5,5- diethybarbituric acid) from Fisher Scientific Co. (Springfield, NJ, U.S.A.); Tricine and Tris from General Biochemicals (Chagrin Falls, OH, U.S.A.); sucrose from Mallinckrodt (Jersey City, NJ, U.S.A.); albumin from Pentex (Kankakee, IL, U.S.A.); cyclic [$^3$H]AMP from Schwartz—Mann (Orangeburg, NY, U.S.A.).

**Preparation Procedure**

Male Sprague-Dawley rats from Camm Research Institute (Wayne, NJ, U.S.A.) weighing 200 g were stunned and decapitated at $4^0$C. Cerebral cortex and corpus striatum from 12 brains were sliced and collected in a volume (4ml per each cortex or striatum slice) of ice-cold 10 mM-Tricine/Tris buffer, pH 7.4, with 2mM-sucrose and homogenized. The homogenates were centrifuged at 2400rev./min for 20min (IEC 809 head) in $5^0$C cold-room. Supernatants discarded, pellets were washed and recentrifuged again by same procedure.

**Preparation of $Ca^{2+}$ treated particles**

The isolated pellets obtained from both corpus striatum and cortex were separately resuspended in buffer with 0.5mM-$CaCl_2$, and used immediately or after storage at $-70^0$C, protein contain was measure [1]. After thawing for 30min at room temperature ($24^0$C), the $Ca^{2+}$-treated adenylate cyclase particulate preparations (Ca2+treatcd particles) were washed once more by using the same centrifugation procedure with 10mM-Tricine/Tris, pH 7.4. The pellets were resuspended in a volume of the same medium needed to dilute the particles and used at the concentration reported under graphics 3 and figure C.

**Preparation of EDTA-treated particles**

Particles were obtained and washed twice in an equal volume of the same solution by using the centrifugation procedure described above for the preparation of $Ca^{2+}$ treated particles. The pellets obtained from cortex were separately resuspended in the same buffer and separately stored at −70°C. The protein contents of these EDTA-treated adenylate cyclase particulate preparations were determined by a Lowry colorimetric method, using a calibration curve of bovine serum albumin, as the standard [1]. The frozen particles were used after thawing for 30min at room temperature (24°C), samples used as described in graphics 1 and 2.

**Assay of Adenylate Cyclase activity**

Unless otherwise indicated, samples of treated particles obtained from 20mg fresh wt. of total cortex tissue were incubated at the legends indicated temperature and time in a reaction mixture (1 ml) containing: 40µmol of Tricine/Tris buffer, pH 7.4, 6.67µmol of caffeine and indicated concentration of MgATP adjusted to final pH7.4 with Tris. Modifications to the assay medium are indicated in the legends. The incubation was terminated by heating in a boiling-water bath for 5 min. Blanks (zero time) were obtained for all assays by incubating a previously boiled particulate preparation under the same conditions. The reaction mixtures were centrifuged in a Sorvall RC2B centrifuge at 5000g for l0min and the supernatants were collected and assayed for cyclic AMP formed. Whenever immediate measurement was not possible, the supernatants were stored at −70°C and later used for their cAMP assay.

**Assay of cyclic AMP**

Cyclic AMP formed by adenylate cyclase activity was measured by the Gilman method [2]. This is a saturation assay procedure [3] using a 0.4ml reaction mixture containing 50μl of the supernatant from the incubation mixtures, were diluted if required, to adjust the content of cyclic AMP to less than 5pmol. The samples and corresponding zero time controls, l00μl of 35mM-sodium acetate buffer, pH4.0, containing 11.4μg of albumin, 50μl of cyclic [$^3$H]AMP ($NH_4^+$ salt, 0.55pmol, containing 30,000c.p.m.), and a mixture of 50μl (8mg/ml) of binding protein. The latter, was prepared from fresh bovine muscle by the Miyamoto's method [12] [4] with 50μl of 5mM-potassium phosphate buffer, pH7.0.

The cyclic AMP assay mixture, was incubated in an ice bath for 1 h, thereafter 1ml of 20mM-potassium phosphate buffer, pH 7.0, was added and the mixture was filtered through a type HA Millipore filter (0.45μm pore size) in a 3025 Millipore Sampling Manifold, to separate the precipitated binding-protein-cyclic[$^3$H]AMP complex. The filters containing the precipitate were washed with 3ml of 5 mM-potassium phosphate buffer, pH6.0, placed in vials with l0ml of Bray's [5] scintillation fluid.

Radioactivity was measured in a Beckman LS-l00 scintillation counter. To obtain standard curves for a calculation of the results, calibration curves of known amounts of cyclic AMP were prepared for every experiment. Results are expressed as nmol of cyclic AMP formed/h for per $Ca^{2+}$-treated particles from 1g fresh wt. of total cortex tissue. Cyclic AMP formed by EDTA-treated particles from brain corpus striatum or cortex was assayed by using 100μl samples of supernatants from the incubation mixtures and adjusting the amounts of the other reagents used in the method described above to correspond to a final volume of 0.3 ml. Results are expressed in nmol of cyclic AMP formed/h per g of protein in the EDTA-treated particles as determined by a Lowry colorimetric method [1], with bovine serum albumin as standard.

**Introduction**

A role for adrenergic signal transmission, could be based on the finding of catecholamines-dependent increase of cyclic AMP concentrations [6] [7] [8] [9], and the latter stimulatory action on the protein kinases found in neural tissues [10].

A role from stimulation of adenylate cyclase leading to memory conformation has been shown in the hippocampus [11]. Other studies indicate that inhibition of adenylate cyclase could also be involved in memory formation [12]. Long-term memory (LTM) may surge from the convergence of signal transduction pathways which increase synaptic efficiency [13].

A neurotransmitter dependent control of adenylate cyclase in relation with other membrane located proteins including its regulatory expression by the system GTP plus G protein is illustrated in fig A to show possible ways of interactions for short term memory. The adenylate cyclase extraction from the membrane as well as other membrane enzymes and proteins, results in the lost of properties related to their physiological function. Some of these properties could be preserved by purifying the membranes, which allows obtaining a system, which could be study isolated from the many others present in isolated tissues.

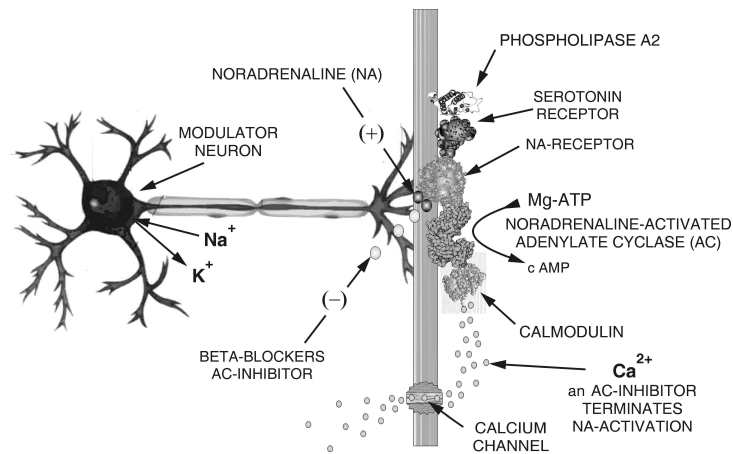

**Fig. A: Illustration of noradrenaline-activation of adenylate cyclase.** The figure illustrates that modulating neurons, unloads the noradrenaline neurotransmitter, in its synapses with a neuron, noradrenaline (epinephrine) binds a receptor associated to G protein that in the presence of GTP promote the neurotransmitter noradrenaline (NA) activation of adenylate cyclase and the transformation of its substrate Mg-ATP, into intracellular cyclic AMP. Presence of beta-blockers prevents physiologic effects of noradrenaline.

The characterizations of particulate preparations allow supporting the findings which indicate that brain controls through catecholamine levels the intracellular cAMP [14]. The latter, modify metabolic interactions, which are illustrated in fig B, and could lead to conform intracellular pathways formative of emotional neuronal circuits, that could modulate emotional perception and memory. The latter integration may lead to overall emotional intelligence [15].

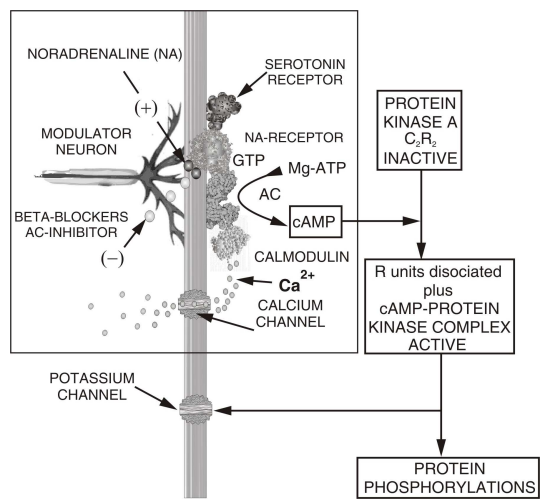

**Fig. B: Cyclic AMP modulator effects**. In humans, cyclic AMP activates protein kinase A (PKA) that has an inactive form consisting of two catalytic and two regulatory units ($C_2R_2$). The catalytic units are block by the R units and cyclic AMP acts by dissociating the regulatory subunits from the catalytic ones, allowing the latter to phosphorylate substrate proteins or enzymes, changing these ones from inactive to active forms or vise verse. The level of cAMP is control by phosphodiestearase breaking down cAMP into AMP. These interactions may result in neuronal pathways leading to changes on a neuron. The latter, could condition its answer to subsequent stimulus, and therefore, allowing to conjecture that these are functional to memory formation.

Acute stress results in a decrease in cyclic AMP content of the hypothalamus and the adenohypophysis. The nature of the stress may have different effects in different areas of the brain; since a decrease in hypothalamic cyclic AMP was observed with an increase in adenohypophysial cyclic AMP [16].

Beta-blockers, by competing with adrenalines, block the molecular response that involves activation by noradrenaline of adenylate cyclase of brain and many other tissues.

In the absence of beta-blockers, adrenalines activation leads to an increment of cyclic intracellular AMP, as illustrated in figure B.

The level of adrenalines became incremented by stress. Prolonged exposition time of some tissues to this hormone, may connect a psychic response, developing into trauma. Hence, allowing the conjecture, that stress trough adrenaline, allows for the etiology of psychosomatic illness. Suggesting, that could be relevant to inquire at the molecular level, for evidence of adrenalines affecting the lack of stability of enzymes.

## Results

It was assumed that the molecular characterization of brain tissue adenylate-cyclase in interaction, with other membrane proteins like the adrenaline receptor and G protein allows in an isolated system, to measure response to the adrenalines, for a biochemical simulation of this enzyme physiological response.

EDTA-treated particles from rat brain and corpus striatum and cortex were used to minimize any contribution by residual metals to the conditions under which the effects of $Mg^{2+}$ and $Ca^{2+}$ on noradrenaline stimulation of adenylate cyclase were assayed (graphics 1 and 2).

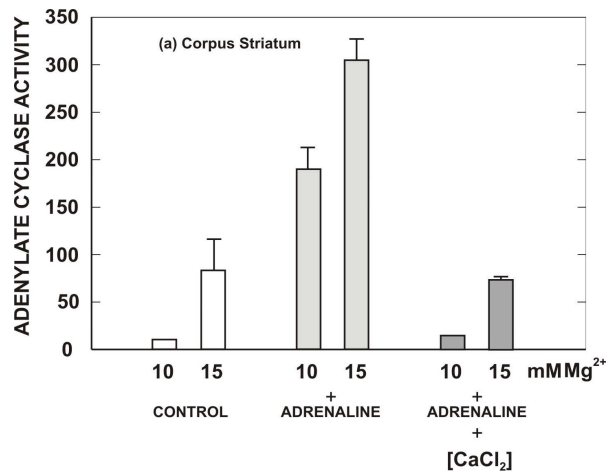

**Graphic 1: Noradrenaline effect on the activation of adenylate cyclase of cerebral corpus striatum.** 76μg of protein of the EDTA-treated-particle preparation from rat brain's isolated membranes of corpus striatum were incubated, in the indicated concentration of 10 mM-$MgCl_2$ and 15 mM-$MgCl_2$, and constant 1mM ATP with 2% GTP at $25^0$C during 60 minutes at pH 7.4; Control (without addition); + adrenaline (+0.1 mM-noradrenaline); + adrenaline + $Ca^{2+}$ (+ 0.1 mM-noradrenaline and 0.3 mM- $CaCl_2$). The cAMP (cyclic AMP) formed expressed by activity units (nmol of cAMP formed per hour/g of proteins in the EDTA particle preparation of the membrane).

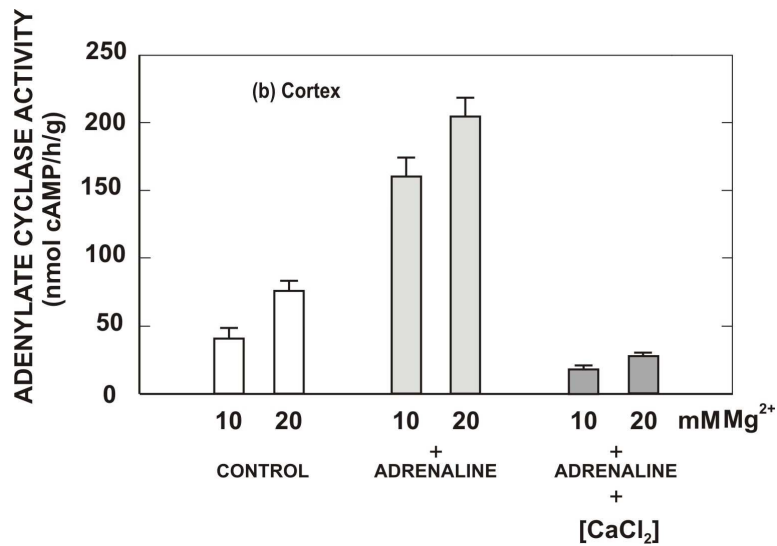

**Graphic 2: Noradrenaline effect on the activation of adenylate-cyclase of cerebral cortex.**
100μg of protein of the EDTA-treated-particle preparation from rat brain's isolated membranes of cortex were incubated in 10 mM-$MgCl_2$ and 20 mM-$MgCl_2$ and constant lmM-ATP with 2% GTP, plus additions indicated on tests at $25^0C$ during 60 minutes. Control (without addition); + adrenaline (+0.1 mM-noradrenaline); + adrenaline + [$Ca^{2+}$] (+ 0.1 mM-noradrenaline and 0.3 mM- $CaCl_2$). The cAMP (cyclic AMP) formed expressed by activity units (nmol of cAMP formed by hour/gr of proteins in the particle preparation of the membrane).

Graphics 1 and 2 show increment of cyclic AMP production as a response to adenylate-cyclase activation by nor-adrenaline, and decrease of cyclic AMP formed by inhibition of the enzyme by the calcium addition.

Brain cells show a higher level of cytoplasm $Mg^{2+}$ and lower level of $Ca^{2+}$ with regard to those found in mitochondria and in the extracellular fluids [17] [18]. Adrenergic synaptic transmission is associated to transient alterations of $Ca^{2+}$ gates and/or release of $Ca^{2+}$ bound to intracellular sites [17] [18].

Nerve cells [19] at the physiological level show transient alterations of $K^+$ flowing to the outside and $Na^+$ inside, changing membrane potentials which are coupled to AC. Studying particulate membranes preparations allows isolation of the neurotransmitter activatory effect, but may require higher $Mg^{2+}$. Graphics 1 and 2 report characteristics of the enzymes, which therefore do not depend on $Na^+$, $K^+$-exchanges, related to transient states of ionic channels and membrane potential.

Turnover requirements suggest that electrochemical transference, allows a maximal total of 10 molecules of water around $K^+$ and 16 around $Na^+$. These, hydration spheres could be transfer to ligand proteins, either to favor the release of the chelating ions: $Mg^+$ and $Ca^+$, or to participate in protein conformational changes, driven by solvatation water.

The thermodynamics of water solvating protein involve about 1 Kcalorie per molecule of water. Hence, the thermodynamic efficiency of physiological coupling is avoided in order to separately study the saturation effect of $Mg^+$ and $Ca^+$, to produces respectively activatory and inhibitory conformational changes on the enzyme-membrane system. Accordingly, the modulatory ranges, for the divalent metals could be expected to require significant lower concentrations at the physiological level, that the observed ones in graphic 1 and 2.

The thermodynamic of a physiological complete turnover cycle, involves an ATP cycle, in which the $ATP^{4-}$ chelates the divalent metal, which is released to the ligand proteins (channels, G-protein, AC), if a couple ATPase generates the weaker chelator $ADP^{3-}$. Multiple couplings, may allow changes in the solvatation states of participating proteins. The driving forces could be complemented by compartmentalized mass-action of water. This one could contribute to release the inhibitory $Ca^{2+}$ from AC, and restore the enzyme as capable to bind a new neurotransmitter molecule. A turnover, implicating solvatation water, may be favored by the strong exergonic reactions between $Ca^{2+}$ and water to generate the hydrations sphere of free $Ca^{2+}$.

Kinetic characterization of adenylate cyclase response to $Mg^{2+}$ and $Ca^{2+}$ as shown in graphics 1 and 2 corresponds with the expected physiological function within an integrated turn-on and turn-off control, by $Mg^{2+}$ and $Ca^{2+}$ fluxes, which may allow short-term memory formation of cAMP for latter metabolic transformation on long term-memory.

Consequently, enzyme activation could be characterized like the molecular answer to the stimuli, mediated by the adrenalines known as "fight or flight", and/or the initial step on the formation of emotional memory [15] [20] [21] [22]. Calcium unloads within the brain cells, to inhibit the enzyme and terminate the activator impulse initiated by the neurotransmitter [23] [24].

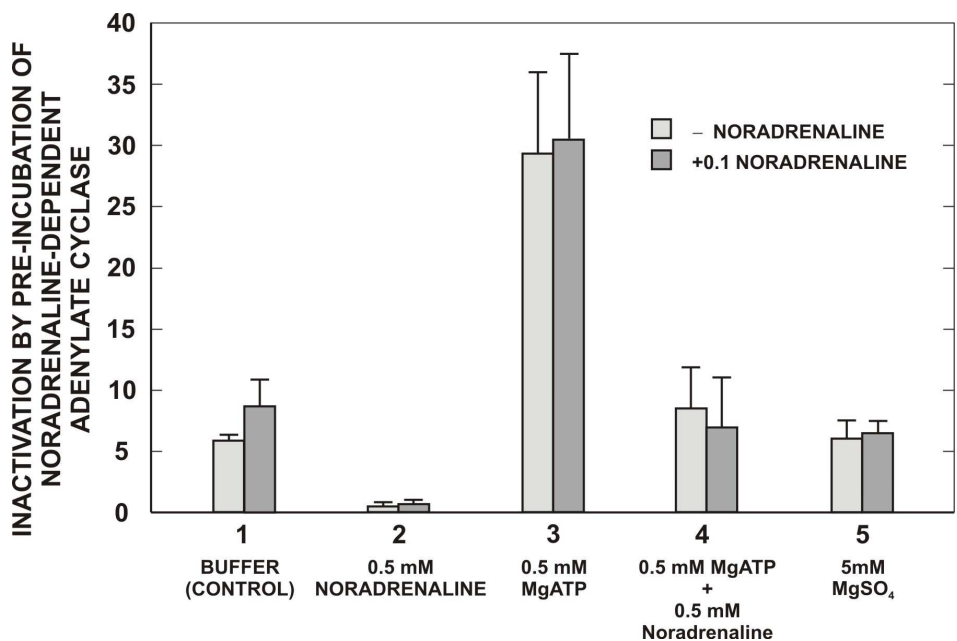

**Graphic 3: Destabilizing effect on adenylate cyclase by overexposure to noradrenaline.** Samples from 36mg fresh tissue wt. of $Ca^{2+}$-treated-particle, from isolated membranes of rat brain cortex in 10mM-Tricine/Tris at pH 7.4, in a final volume of 0.9ml, and the additions indicated in the figure from 1 to 5, were preincubated at 38°C during 3 hours, as an assay of enzyme stability. The pre-incubation mixtures were centrifuged at l700rev/min from l5min in an IEC model 809 head at 20°C. The pellets were washed once by the procedure indicated in the Experimental section, with 3ml of the same buffer, and suspended in 0.36ml of the same. Samples of pre-incubated particles (from l5mg fresh wt. of tissue) were incubated in a final volume of 0.5ml. The residual adenylate-cyclase activity was measured in 1 mM ATP free of GTP in the absence of noradrenaline or basal conditions (1 to 5-ligth bars) and with addition 0.1mM noradrenaline (1 to 5 dark bars).

The experimental approach reported in graphic 3 was intended as a test tube simulation of prolonged stress by studying the stability of different complex forms of the enzyme to prolonged exposure to norepinephrine. The test would be revealing that the conformation of the complex enzyme neurotransmitter is more sensitivity to the inactivating effect of body temperature (37~38$^o$C).

$Ca^{2+}$-treated particles were used to simulate a possible residual $Ca^{2+}$-dependent refractory period. In other words the treatment allows departing from an inhibited state of the enzyme, which could have masked the inactivatory effects by the neurotransmitter itself.

Also, the activity of enzyme was measure by incubation in ATP free of GTP, in order to avoid activatory effects of NA, masking it's over exposure-dependent inhibitory effects. The results show that the enzymatic activity has notably decrease during the preincubation period with noradrenaline (assay **2**: 0.5mM Noradrenaline).

Hence, it could be inferred that when the enzyme-membrane was over exposed to noradrenaline for a long period of time, the adrenaline-enzyme complex acquired a more unstable state than the uncomplex enzyme used as control (assay **1**: Buffer/control).

The preincubation of the enzyme-membrane, with a saturating substrate concentration, measures protection on the active and regulatory center of adenylate cyclase, by formation of the enzyme-substrate-complex (assay **3**: 5mM Mg ATP). The preincubation induced 300% activation by formation of the substrate MgATP-enzyme-complex indicating that the membrane stabilized a temperature-induced more active enzyme conformation. However, this treatment protected the active site of the enzyme, but uncoupled any residual lack of GTP dependent activatory effect of noradrenaline.

Graphic 3, also shows that if the enzyme had been preincubated with buffer plus $Mg^{2+}$ (**5**: 5 mM $MgSO_4$) the enzyme keeps considerable activity but, also shows the uncoupling response.

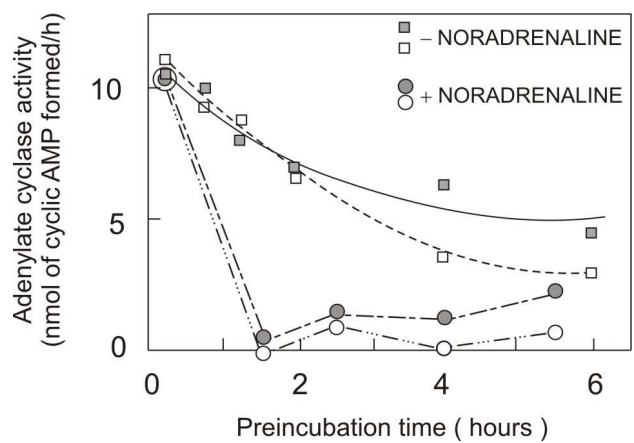

**Fig. C. The noradrenaline effect on the stability curves of adenylate cyclase.** Samples of a $Ca^{2+}$-treated particle preparation of brain cortex containing 100 mg of protein, were preincubated at 38$^0$C in 6.25 ml of 6 mM-Tricine/Tris buffer, pH 7.4, in the absence: (□, ■) and in the presence (○, ●) of 0.2 mM noradrenaline. Samples were taken at the indicated times and centrifuged at 5000 g for 5 minutes in a Sorvall RC2B centrifuge. The pellets were resuspended in 10 mM-Tricine/Tris buffer, pH 7.4. Samples containing

2.4 mg of these washed particles were incubated in a final 0.5 ml of the standard reaction mixture with ATP (without GTP) at $15^0$C in the absence (□,○) or in the presence (■,●) of 0.1 mM-noradrenaline and adenylate cyclase activity was determined and reported as adenylate cyclase activity (nmol of cyclic AMP formed/h per particles yielded from 1 g of fresh wt. of tissue).

The condition 2 in graphic 3 was reproduced at lower noradrenaline concentration in the preincubation curves shown in fig C. The one tested for residual noradrenaline activation and the one for basal activity of the enzyme, show similar rapid inactivation in less than 1.5 hours.

A the cellular level, a cascade of molecular events induced by increment of cAMP as illustrated in graphics 1 and 2, results in changing the state of activity of the protein kinases. Hence, with intensification of nervous cells connectivity and allowing sustenance of long term memory [15]. A cellular decrease of cAMP levels (stress) could therefore have opposite effects.

## Discussion

**Metabolite-modulation may allow ionic control of the responsiveness to hormones of adenylate cyclase activity**

Atkinson [25] formulated energy charge: $E_C = \frac{1/2[ADP]+[ATP]}{[AMP]+[ADP]+[ATP]}$, which evaluates the metabolic state of cells, assumes that when all the adenylate nucleotide is in the form of AMP, the $E_C$ value equals 0. Hence, when all is present as ATP, $E_C$ equals 1. At high $E_C$ values, the ATP generating reactions are inhibited, and the ATP consuming ones became stimulated.

However, this parameter ignores the chelating order: $ATP^{4-} \gg ADP^{3-} > AMP^{2-}$. This sequence reveals an ionic control, because generation of $ATP^{4-}$ and other chelating metabolites decreases the concentration of free ions like $Mg^{2+}$ [26]. The latter, is required as shown in graphics 1 and 2, to allow the activation of adenylate cyclase by hormones or neurotransmitters, as illustrated in the figure C.

The emotional pathway, in the reaction of stress, stimulates the adrenal glandule to produce adrenaline in blood, activating the adenylate cyclase in tissues like fat tissue [26] [27], etc. The adenylate cyclase of fat tissue (EC 4.6.1.1; ATP pyrophosphate-lyase (cyclizing), responds to the adrenaline stimuli as a function of metabolite-modulated interactions, between active centers for substrate Mg-ATP complex, to produce cAMP (cyclic AMP) . The regulatory controls of the enzymatic activity and its answer to adrenaline, it's exerted by free $Mg^{2+}$ versus free $ATP^{4-}$ [26] [27]. Their equilibrium allows modulator effects, in a receptor site for $Mg^{2+}$, acting like a sensor of energetic availability [23] [24] [28] [29].

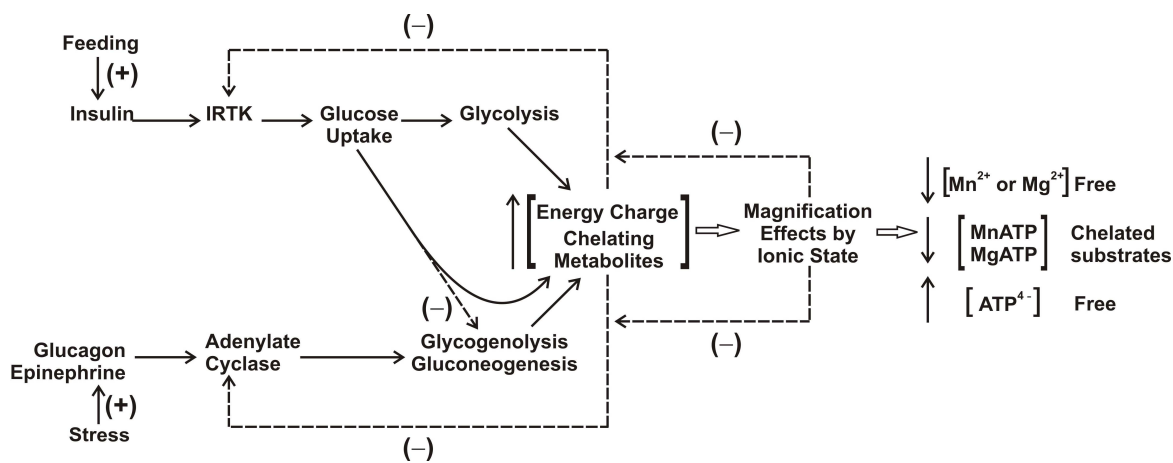

**Fig. D. Integration of the ionic and metabolic cellular controls of the insulin receptor tyrosine kinase (IRTK) and adenylate cyclase activities (AC).** An increment in the concentration of Chelating Metabolites [↑] produces the release of free $ATP^{4-}$, a negative modulator of AC and IRTK. Magnification effects by simultaneous decrease [↓] of the chelated substrates: concentrations of MnATP and MgATP, and of the activatory modulators the free ions: $Mn^{2+}$ and $Mg^{2+}$.

Excess of $ATP^{4-}$ or $citrate^{-1}$, indicates that the intracellular level of energy reserves are high and, in the case of the fat cells, saturation of the $ATP^{4-}$ site, inhibits the stimulatory effect of the lipolytic hormones: ACTH and the stress hormones glucagon and l-adrenaline, by decreasing their affinity for the enzyme [26] [27]. Inactivation of the adenylate cyclase shifts the cells metabolic response to triglycerides synthesis. Metabolic-dependent changes in the concentration of free ATP versus chelated ATP could be amplified, by changing the ionic equilibrium as shown of in figure C. The latter, exemplifies an integrative metabolic control of the hormonal or neurotransmitter responsiveness of cellular adenylate cyclase.

An excess of ionic $Mg^{2+}$, results from a low level of the chelating metabolites which are intermediates of glucolysis. Saturation by free $Mg^{2+}$, of the regulatory site magnifies the affinity of the enzyme for the stress hormones [24] [28] [29].

In terms of the physiological role of the observed preincubation relationships (graphic 3), exposed to buffer alone, would correspond with starvation, decreasing intracellular ATP. Therefore, nutrition deficiency could be equated with lack of ATP, which could decrease resistance to stress.

In children's development, could be predicted from graphic 3, assay **2,** that alimentary insufficiency, or lack ATP could expose AC to a greater tendency to stress-inactivation and may affect learning link with emotional memory [15] [20] [21] [22]. This effect has great social impact, because under nutrition may decrease the children efficiency at learning, and it would affect cerebral development, leading to lower their intellectual coefficients.

Beta-blockers mechanism, like propanolol action in the pharmacologic receptor for noradrenaline, can be explained by its antagonistic binding to the adrenaline receptor, which controls the activity of the adenylate-cyclase enzyme [23] [24] [28] [29]. Therefore, it may interrupt the cellular activity which controls consolidation of memory pathways [15].

Beta-blockers are used to protect the vascular system from the effects of adrenalines, when antagonizing the adrenaline binding to adenylate cyclase, mitigate or attenuate the effects on cells of the circulating adrenalines in blood, or in the cerebral spinal fluid. Nervous system peptides mimic the effect of opiates [30] (Hughes J. 1975 Brain Res. 88,295-308.). Narcotics transient inhibition and late positive regulation of neural circuits involve adenylate cyclase hours after the initial event (Schaefer) [31].

Beta-blockers could be used in that context, since more recently have been shown to have the therapeutic effect of protecting people, from the emotional impact associated with formation of harmful memories and post traumatic stress [21] [32] [30] [33].

There is a short term memory, an intermediate memory and a long term memory. The beta-blocker, if given before formation of long term memory, could diminish the intensity of the association between perception of the event and traumatic bodily response, like tachycardia, etc. Therefore, allowing a scenario of turning emotional memory, into a factual memory [30] [32].

These studies, agree on that, in the process of memory affirmation, there are two different phases: the cell mechanism that establish the changes in the form of neuronal response previously described [14]. Plus, connections and fixation, which reorganize and restructure the neuronal circuits, allowing to recuperate and store memories, these may involve cyclic AMP in the induction mechanisms of several enzymes [34] [35] [36] [37] [38].

Then, the noradrenaline activated pathway [14] and the mediated by acetylcholine, are differentiable systems of memory, functionally separated, but integrated for conscious memory capability. Thus, the emotional pathway configures the possibility of acquiring skills, habits, affects and aversions, etc. [15] [20] [21] [22]. The spinal-brain barrier separates blood from cerebral spinal fluid. The locus cereleus is rich in neurons uploading nor-adrenaline at the synapse, and it's been related with conducts disorders, like panic, avoidance, etc [32] [30] [33].

Therefore, the effect of noradrenaline on the control of adenylate cyclase activity, and the stability of this labile enzyme [7] [14], may have significance in the aetiology of stress-related diseases.

Thus stress, through the destabilizing and inhibitor effect of these hormones on AC modifies the enzyme itself, and its ability to control brain cells activity and feedback response. This, leads to pathologies of psychosomatic etiology.

**Can metabolites by regulating the activity and stability of the adrenaline response to AC modulate animal longevity?**

The study of longevity consistently shows that over-alimentation decreases in more than half, the average of life expectancy of the studied animals. Sub-alimentation prolongs the life expectancy, with respect to control animals [20] [30].

Genetic disruption of AC5 increases mouse life span and confers resistance to aging-related conditions, including bone loss and cardiomyopathies [39]. It is proposed that these beneficial effects may be the result of the increased activity of second messenger signaling proteins [39].

The enzymes adenylate cyclase and insulin receptor tyrosine-kinase are characterized as being inhibited by $ATP^{4-}$ concentration in excess of free $Mg^{2+}$. At similar concentration, the chelating interaction of ATP with $Mg^{2+}$, forms MgATP allowing little dissociation, and therefore decreasing the ionic form ($Mg^{2+}$), suppressing the activator effect of these hormones in the enzymes. The condition of $ATP^{4-}$ in excess reinforces inhibition by absence of $Mg^{2+}$. The latter relationship, would be present in over-fed animals, and would decrease the response to stress, a possible explanation of over-eating as a defense mechanism to emotional stress. On the other hand, tendency to sub-alimentation would be favoring longevity [40] [41].

**Conclusions**

Enzymes maintain cell levels by continuous synthesis, opposed to the inactivation, in certain cases a thermal phenomenon. In the test tube, acceleration of the thermal effect, by the stress hormone adrenaline, is not compensated by its synthesis, and the enzyme level would show more clearly, the accelerated tendency to decrease activity.

In the assay tube, the enzyme response to became activated by noradrenaline, could correspond to a physiologically function in pathway of emotional memory formation. The decrease of its activity by blockers may alter the link between emotion and memory. The mind is considered responsible, through maintaining the increment of adrenaline hormones, during an extended period of stress, for producing somatic illnesses, which not only affects the brain, but alternatively others organs.

Noradrenaline forms with the enzyme a complex, which in addition to activate the enzyme, induces a thermal labile state accelerating loss of its activity. This effect could be proposed as the molecular mechanisms of the physiological responses to stress, would be equivalent to a molecular connection between mind and body, or in other words, mind inducing higher levels of noradrenaline, over long period of time affect the enzyme which controls in large measure intracellular, process describable as somatic trauma.

The observed progress of psychosomatics illnesses shows at first a hipper functionability followed by a decrease function, this would be equivalent to anxiety followed by depression. Analogous evolution of other disorders, affecting other tissues like thyroid, would be a hipper thyroid stimulation followed by hippo thyroid function.

The biochemical systems involved in decreasing enzyme concentrations are poorly understood at the present, especially those that affect the turnover of membrane components. This connection would function by stress response, increasing secretion of adrenaline, initially stimulating the enzyme, but latter destroying the enzyme. Hence, prolonged expositions of the enzyme to stress hormones, in excess of the cell capacity to restore it, may modify the cell's metabolism and the induction mechanism of some enzymes [34] [35].

Molecular level studies are concordant with those at the organism level, since they demonstrate that, if adrenaline is blocked by propranolol, it has a psychotherapeutic activity. The effects of Beta-Blockers can be explained at molecular level, because by antagonizing the union between noradrenaline and its associated receptor on the enzyme membrane the active center of the latter becomes protected.

In this way, beta-blockers may lessen emotional consequences dependent on adrenaline increase during traumatic experiences [21] [33]. In the absence of blockers, the stress effect prolongs the adrenaline action, even further than its physiological activatory function. This overexposure would causes dysfunctions in capabilities like thinking, planning, judgment and memory. Consequently, beta-blockers administration would be efficient by preventive effects of stress rather than showing a curative action.

The contributions of Brydon-Golz, S., Ohanian H. and Harris, R. are acknowledged.